\documentclass[pdflatex,sn-mathphys-num]{sn-jnl}

\usepackage{graphicx}%
\usepackage{multirow}%
\usepackage{amsmath,amssymb,amsfonts}%
\usepackage{amsthm}%
\usepackage{mathrsfs}%
\usepackage[title]{appendix}%
\usepackage{xcolor}%
\usepackage{textcomp}%
\usepackage{manyfoot}%
\usepackage{booktabs}%
\usepackage{algorithm}%
\usepackage{algorithmicx}%
\usepackage{algpseudocode}%
\usepackage{listings}%

\theoremstyle{thmstyleone}%
\theoremstyle{thmstyletwo}%

\theoremstyle{thmstylethree}%

\begin{document}

\title[Article Title]{\textbf{ Comparative Analysis of Deep Learning Architectures for Breast Region Segmentation with a Novel Breast Boundary Proposal}}


\author*[ 1,2,3]{\fnm{Sam} \sur{Narimani}}\email{sam.narimania@gmail.com }

\author[ 3,4]{\fnm{Solveig} \sur{Roth Hoff}}\email{solveig.roth.hoff@ntnu.no}


\author[ 5,6]{\fnm{Kathinka} \sur{Dæhli Kurz}}\email{kathinka.kurz@uis.no}

\author[ 7,8]{\fnm{Kjell-Inge} \sur{Gjesdal}}\email{kjell.inge.gjesdal@nordiccad.com}

\author[ 9,10]{\fnm{Jurgen} \sur{Geisler}}\email{jurgen.geisler@medisin.uio.no}

\author[ 1,2,3]{\fnm{Endre} \sur{Grovik}}\email{endre.grovik@gmail.com}

\affil[1]{\small{\orgdiv{ Department of Physics}, \orgname{Norwegian University of Science and Technology}, \orgaddress{ \city{Trondheim},  \country{Norway}}}}

\affil[2]{\small{\orgdiv{ Research and Development Department}, \orgname{More og Romsdal Hospital Trust}, \orgaddress{ \city{Aalesund},  \country{Norway}}}}

\affil[3]{\small{\orgdiv{ Department of Radiology}, \orgname{More og Romsdal Hospital Trust}, \orgaddress{ \city{Aalesund},  \country{Norway}}}}

\affil[4]{\small{\orgdiv{ Department of Circulation and Medical Imaging}, \orgname{Norwegian University of Science and Technology}, \orgaddress{ \city{Trondheim},  \country{Norway}}}}

\affil[5]  {\small {\orgdiv{ Department of Electrical Engineering and Computer Science}, \orgname{The University of Stavanger}, \orgaddress{ \city{Stavanger},  \country{Norway}}}}

\affil[6]  {\small {\orgdiv{ Department of Radiology}, \orgname{Stavanger University Hospital}, \orgaddress{ \city{Stavanger},  \country{Norway}}}}

\affil[7] {\small  \orgdiv{ Department of Diagnostic Imaging}, \orgname{Akershus University Hospital}, \orgaddress{ \city{Lorenskog},  \country{Norway}}}

\affil[8] {\small \orgname{ NordicCAD AS}, \orgaddress{ \city{Aalesund},  \country{Norway}}}

\affil[9] {\small  \orgdiv{ Institute of Clinical Medicine}, \orgname{University of Oslo}, \orgaddress{ \city{Lorenskog},  \country{Norway}}}

\affil[10] {\small  \orgdiv{ Department of Oncology}, \orgname{Akershus University Hospital}, \orgaddress{ \city{Lorenskog},  \country{Norway}}}




\abstract{\textbf{Purpose:} Segmentation of the breast region in dynamic contrast-enhanced magnetic resonance imaging (DCE-MRI) is essential for the automatic measurement of breast density and the quantitative analysis of imaging findings. This study aims to compare various deep learning methods to enhance whole breast segmentation and reduce computational costs as well as environmental effect for future research.

\textbf{Methods:} We collected fifty-nine DCE-MRI scans from Stavanger University Hospital and, after preprocessing, analyzed fifty-eight scans. The preprocessing steps involved standardizing imaging protocols and resampling slices to ensure consistent volume across all patients. Using our novel approach, we defined new breast boundaries and generated corresponding segmentation masks. We evaluated seven deep learning models for segmentation namely UNet, UNet++, DenseNet, FCNResNet50, FCNResNet101, DeepLabv3ResNet50, and DeepLabv3ResNet101. To ensure robust model validation, we employed 10-fold cross-validation, dividing the dataset into ten subsets, training on nine, and validating on the remaining one, rotating this process to use all subsets for validation.
 
\textbf{Results:} The models demonstrated significant potential across multiple metrics. UNet++ achieved the highest performance in Dice score, while UNet excelled in validation and generalizability. FCNResNet50, notable for its lower carbon footprint and reasonable inference time, emerged as a robust model following UNet++. In boundary detection, both UNet and UNet++ outperformed other models, with DeepLabv3ResNet also delivering competitive results. }





\maketitle

\section{Introduction}\label{sec1}

Whole breast segmentation is a pivotal step in assessing the risk of breast cancer \cite{acciavatti2023beyond}. Accurate breast region segmentation can be beneficial not only for lowering computational cost for predicting breast cancer but also to focus on quantitative analysis of breast cancer \cite{lew2024publicly,ojala2001accurate}. Among all medical imaging modalities, MRI plays a significant role in high quality visualization of the whole breast \cite{jiao2020deep}. However, interpretation of breast MRI can be challenging due to noise and artifacts from surrounding anatomical structures \cite{ivanovska2019deep}. 

In recent years, deep learning (DL) techniques have emerged as powerful tools in medical image analysis, offering the potential to automate and improve various tasks such as segmentation \cite{huo2021segmentation,van2020volumetric,yue2022deep}. Segmentation, the process of partitioning an image into multiple regions or segments, is a fundamental step in medical image analysis. In fact, it enables a fast and automatic delineation of Region of Interest (ROI), such as tumors or anatomical regions, with high precision and accuracy \cite{giannini2010fully}. Segmentation of the breast region in MRI images makes it possible to create automatic models for breast density measurement \cite{saffari2020fully}. This process not only enhances the efficiency of data processing but also contributes to the rapid training and analysis of AI models, promoting more environmentally sustainable data processing. Furthermore, precise segmentation of the breast region facilitates the localization and characterization of abnormalities, thereby assisting radiologists in their diagnostic decision-making process \cite{ertacs2008breast}.

In the last decade, significant advancements have been made in breast region segmentation due to the development of numerous AI architectures \cite{zhang2019automatic,piantadosi2018breast,xu2018breast}.
These advancements mark a notable shift in segmentation techniques, transitioning from traditional feature-based machine learning methods, such as clustering \cite{yao2009breast, nie2008development}, to more advanced deep learning approaches, including UNet and its variants \cite{sui2021breast}. These efforts have led to improved diagnostics and more precise stratification of breast cancer tumors \cite{radak2023machine}. Despite these advances, no comparative study has been conducted to evaluate the performance of well-known DL methods for breast region segmentation. Therefore, our study aims to fill this gap by comparing seven prominent DL architectures for segmenting breast regions in DCE-MRI. The goal is to identify the most competitive network that minimizes computational costs while effectively eliminating background noise.

\section{Materials and Methods}\label{sec2}

\subsection{Data}\label{subsec2}

\subsubsection{Data description}\label{subsubsec2}

The dataset utilized in this study consists of DCE-MRI scans obtained from 59 patients at Stavanger University Hospital in 2008. The DCE sequence comprises one pre- and five post-contrast image series with a temporal resolution of 63 seconds. Table \ref{tabel1} provides a detailed description of the dataset and screening parameters.

\subsubsection{Image acquisition}\label{subsubsec2}
All DCE-MRI scans were acquired using a 1.5 Tesla MRI scanner, Philips Intera, with a dedicated breast coil equipped with SENSE technology for high-quality and high-resolution images.
Imaging parameters included T1 weighted fast spoiled gradient echo (FSPGR) sequence, with a scan resolution of 0.9659 x 0.9659 mm\textsuperscript{2} and dynamic acquisition time of 6 minutes and 20 seconds following contrast agent administration.

\begin{table}[h]
\centering
\caption{\centering Detailed Specifications and Imaging Features of MRI Scans}\label{tabel1}%
\begin{tabular}{@{}lll@{}}
\toprule
\textbf{Category} & \textbf{Attribute} & \textbf{Description} \\
\midrule
 & Patient number & 59 \\
\textbf{Study Information}& Weight (kg) & 70.6 ± 8.4 \\
& Patient Position & Head First Prone (HFP) \\
& Number of Images per Patient & 6 (1 pre-contrast, 5 post-contrast) \\
\midrule
 & Scanner Model & Philips Intera MRI Scanner \\
\textbf{Scanner Properties}& Magnetic Field Strength (T) & 1.5 \\
& Coil Technology & SENSE Technology \\
\midrule
& Image Dimensions & (352,352,150), (352,352,140), (352,352,120) \\
\textbf{Image characteristics}& Pixel Spacing (mm) & 0.9659 x 0.9659 \\
& Slice Thickness (mm) & 2 \\
& Field of View (FOV) (mm) & 400 \\
\midrule
 & MRI Sequence & T1 weighted fast spoiled gradient echo (FSPGR) \\
\textbf{Imaging Features}& Repetition Time (TR) & 6.91 \\
& Echo Time (TE) & 3.39 \\
& Flip Angle & 12 \\
\bottomrule
\end{tabular}
\end{table}

\subsection{Preprocessing}\label{subsec2}

The initial dataset, comprised of imaging data in the DICOM standard format, underwent a meticulous cleansing process to ensure data integrity. Subsequently, both pre- and post-contrast images were automatically identified and converted to the NIFTI format, a prerequisite for our modeling endeavors. To ensure consistent data volume, a random oversampling method was applied to minority volumes. This approach simplifies data preparation, making it easier to process before feeding it into the model for further analysis. In addition, breast regions were annotated in detail, adhering to predefined boundary criteria outlined in breast boundary assumptions thereby providing insights for subsequent analyses.

\subsection{Deep learning networks}\label{subsec2}

Over the past few decades, numerous segmentation models have been introduced by researchers. Among these, encoder-decoder based models with skip connections have garnered significant attention due to their effectiveness in retaining important features during training \cite{ronneberger2015u}. In this study, we employed seven widely recognized segmentation architectures—UNet, UNet++, DenseNet, FCNResNet50, FCNResNet101, DeepLabv3ResNet50, and DeepLabv3ResNet101—to train on our dataset. These models were selected for their proven efficacy in medical image segmentation tasks \cite{ronneberger2015u,zhou2018unet++,huang2017densely,long2015fully,he2016deep,chen2017rethinking}.

UNet, introduced by Ronneberger et al. in 2015 \cite{ronneberger2015u}, is one of the most popular segmentation methods. It consists of contraction and expansion pathways connected by skip connections. These skip connections help the model retain important features that might otherwise be forgotten during the training process.
UNet++ is an improved version of UNet, designed to achieve superior results. In UNet++, the skip connections were redesigned to reduce the loss of important features between the contraction and expansion pathways, enhancing the overall performance of the model \cite{zhou2018unet++}.
DenseNet, another architecture utilized in this study, has demonstrated promise in propagating features throughout the model. In DenseNet, every layer is connected to other layers, thereby enhancing feature propagation across the entire network and improving the model's ability to learn complex patterns \cite{huang2017densely}. Given that DenseNet is primarily used for classification tasks, we employed its feature extraction part along with a decoder, excluding skip connections, to examine the impact of their absence in a deeper model.
Next network is FCNResNet comprising a ResNet as the feature extractor and an FCN header \cite{long2015fully} for upsampling or decoding. ResNet's structure, which includes residual blocks, has proven effective \cite{he2016deep}, while the FCN header connects to each feature level, serving as a skip connection.
Last architecture, DeepLabv3 is renowned for its Atrous Spatial Pyramid Pooling (ASPP) block \cite{chen2017rethinking}. Following the ResNet feature extractor, ASPP is applied and subsequently added to the decoder part of the architecture for upsampling.
Table \ref{tabel2} provides practical information about the networks, including learning parameters, the number of layers, and their distinctive features. This comparative analysis offers valuable insights into the strengths and applications of each model in medical image segmentation tasks.

\begin{table}[h!]
\centering
\caption{\centering Model specification and features}\label{tabel2}%
\begin{tabular}{@{}lccl@{}}
\hline
\textbf{Architecture name} & \textbf{Layers} & \textbf{Learning Parameters} & \textbf{Special Features} \\
\hline
UNet & 141 & 31,112,641 & Simple skip connection \\
UNet++ & 240 & 9,119,044 & Dense skip connection \\
DenseNet & 1216 & 70,536,843 & Reusing Feature-maps in subsequent blocks \\
FCNResNet50 & 157 & 32,943,617 & Strong feature extractor alongside FCN header \\
FCNResNet101 & 293 & 51,935,745 & \\
DeepLabv3ResNet50 & 184 & 39,630,593 & Atrous Spatial Pyramid Pooling (ASPP) \\
DeepLabv3ResNet101 & 320 & 58,622,721 &  \\
\hline
\end{tabular}
\label{tab:model-specification}
\end{table}

\subsection{Evaluation}\label{subsec2}
In the evaluation section, we assess the performance of our models using the Dice loss function \cite{dice1945measures} and k-fold cross-validation \cite{kohavi1995study}. The Dice loss function, a promising evaluator for measuring overlap in segmentation tasks, ensures precise model predictions. We employed 10-fold cross-validation meaning that partitioning the dataset into 10 equal parts to validate the model's consistency and performance across different data subsets. The dice loss function was calculated as relation \ref{equation1}, where P and G refer to predicted and ground truth.

\begin{equation}
\text{Dice loss (P,G)} = 1 - 2 \cdot\frac{ P \cap G}{P + G}\label{equation1}
\end{equation}

\subsection{Breast boundaries}\label{subsec2}

Accurately annotating the boundaries of the breast region has been a persistent challenge, as highlighted in previous studies \cite{rosado2015automated}. This challenge arises from the similar intensities observed in imaging for the upper chest wall, fibroglandular tissue, and pectoral muscles, making differentiation difficult \cite{fooladivanda2017breast}.

To improve breast region segmentation, we propose a novel boundary framework designed to exclude high-intensity pixels, such as those from blood vessels near the heart, which are often misidentified by models. The anterior boundary is set at the skin line to remove low-intensity and noisy pixels anterior to it, while the posterior boundary is aligned with the lungs to eliminate low-intensity and noisy pixels dorsal to the chest wall. Additionally, our method incorporates the pectoralis and intercostal muscles, as well as the ribs, to ensure accurate staging of tumors that invade the chest wall.

\begin{figure}[h!]
\includegraphics[width=1\textwidth]{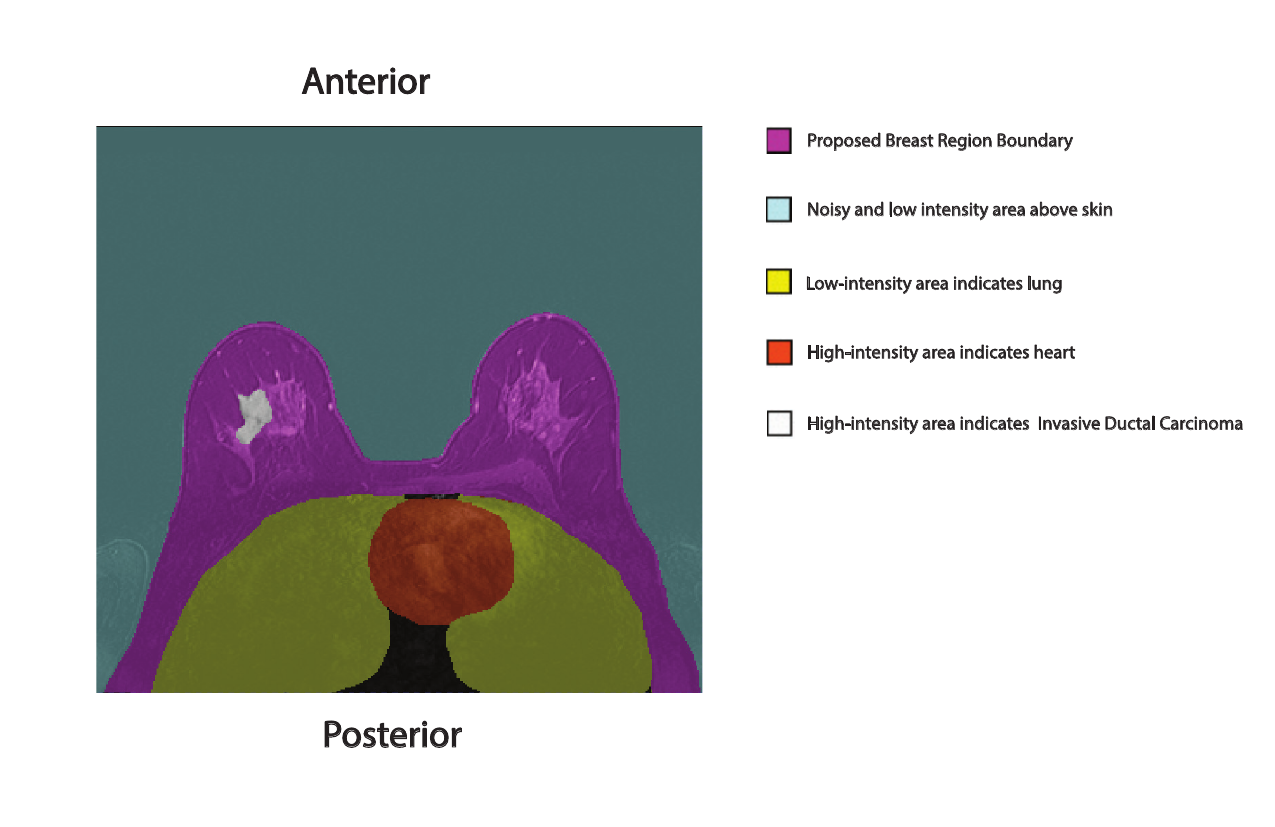}
\caption{\centering Illustration of distinct regions, highlighting the delineation of the proposed breast boundary.}\label{fig-boundary}
\end{figure}

Figure \ref{fig-boundary} illustrates the various regions of interest, highlighting low-intensity areas such as the background and lungs, and high-intensity areas like the heart and lesions, along with the delineated boundaries of the proposed breast region.

\begin{figure}[h!]
\centering
\includegraphics[width=1\textwidth]{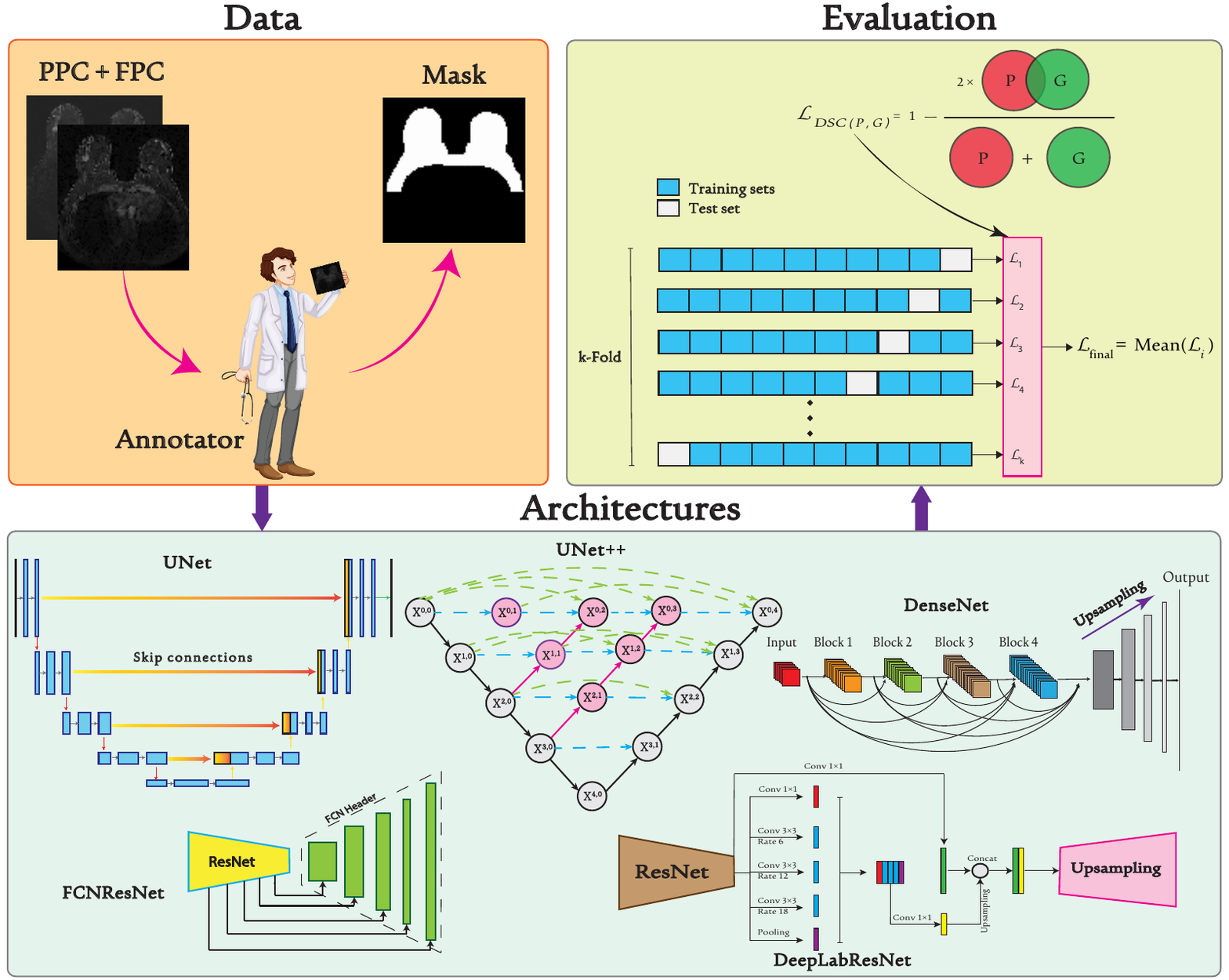}
\caption{\centering Schematic diagram of the study pipeline. Data part depicts data inputs (Pre-Post Contrast (PPC) and First Post Contrast (FPC)), annotator and mask file, and the Architecture block consisting of seven distinct models, each trained individually. Evaluation methods, including 10-fold cross-validation and the Dice loss function, are also illustrated. (Annotator components adapted from Vecteezy.com)}\label{fig1}
\end{figure}

\subsection{Setup}\label{subsec2}

The simulations were conducted on a high-performance computing setup to ensure the efficient training and evaluation of the deep learning models. The hardware configuration utilized in this study includes an AMD Ryzen 9 7900X 12-Core Processor CPU, 32 GB of RAM, and an NVIDIA GeForce RTX 4090 GPU with 24 GB of GDDR6X memory. The power consumption of the GPU is reported to be 450 W by NVIDIA \cite{nvidia4090}, and the entire system consumes roughly 1 kWh of energy during each simulation.

Figure \ref{fig1} depicts different pathways in data processing, the representation of various architectures, and the evaluation metrics and methods employed in our approach. 

\section{Results}\label{sec3}
\subsection{Experiments}\label{sec3}
The input data consisted of both pre- and first post-contrast images, with corresponding masks serving as the ground truth outputs. Training involved various deep learning models using 10-fold cross-validation by utilizing Dice loss function. RAdam optimizer with an initial learning rate of 0.0001 in conjunction with a ReduceLROnPlateau scheduler was utilized to enhance model convergence and performance. This scheduler dynamically adjusted the learning rate based on validation performance metrics, aiming not only to minimize the Dice loss function, but also to accelerate training performance.
Across all models, a consistent batch size of 8 was employed during training, with data shuffled to ensure robust model learning. Finally, a test subset comprising data from two patients was randomly partitioned to evaluate the model's performance on previously unseen data.

\subsection{Model performance and generalizability}\label{sec3}

Table \ref{tabel3} displays Dice training and validation losses across different deep learning architectures at their best epochs. UNet++ achieves the lowest training loss of 0.0112 ± 0.0022, while FCN with ResNet50 also performs well with a Dice training loss of 0.0126 ± 0.0028. On the other hand, UNet architecture stands out for its superior validation results, indicating strong generalization to unseen data essential for real-world applications with validation loss of 0.0448 ± 0.0077. Following closely, UNet++ demonstrates competitive validation performance with losses of 0.0466 ± 0.0167, emphasizing its balanced model performance and generalizability.
In contrast, DenseNet exhibits some of the poorest performance metrics, both in terms of training and validation loss, despite its deeper architecture. On the other hand, DeepLabv3 with a ResNet101 backbone achieves superior validation loss, second only to UNet.

\begin{table}[ht]
\centering
\caption{\centering Training and validation Dice loss for various DL models for $k$-fold cross-validation.}\label{tabel3}
\begin{tabular}{lcc}
\toprule
\textbf{Models} & \textbf{Dice Training Loss} & \textbf{Dice Validation Loss} \\
\midrule
UNet & 0.0146 $\pm$ 0.0024 & 0.0448 $\pm$ 0.0077 \\
UNet++ & 0.0112 $\pm$ 0.0022 & 0.0466 $\pm$ 0.0167 \\
DenseNet  & 0.0163 $\pm$ 0.0038 & 0.0525 $\pm$ 0.0082 \\
FCNResNet50 & 0.0126 $\pm$ 0.0028 & 0.0474 $\pm$ 0.0100 \\
FCNResNet101 & 0.0134 $\pm$ 0.0043 & 0.0497 $\pm$ 0.0067 \\
DeepLabv3ResNet50 & 0.0140 $\pm$ 0.0036 & 0.0469 $\pm$ 0.0085 \\
DeepLabv3ResNet101 & 0.0131 $\pm$ 0.0018 & 0.0462 $\pm$ 0.0034 \\

\bottomrule
\end{tabular}
\label{tab:dice_losses}
\end{table}

\begin{figure}[h!]
\centering
\includegraphics[width=1\textwidth]{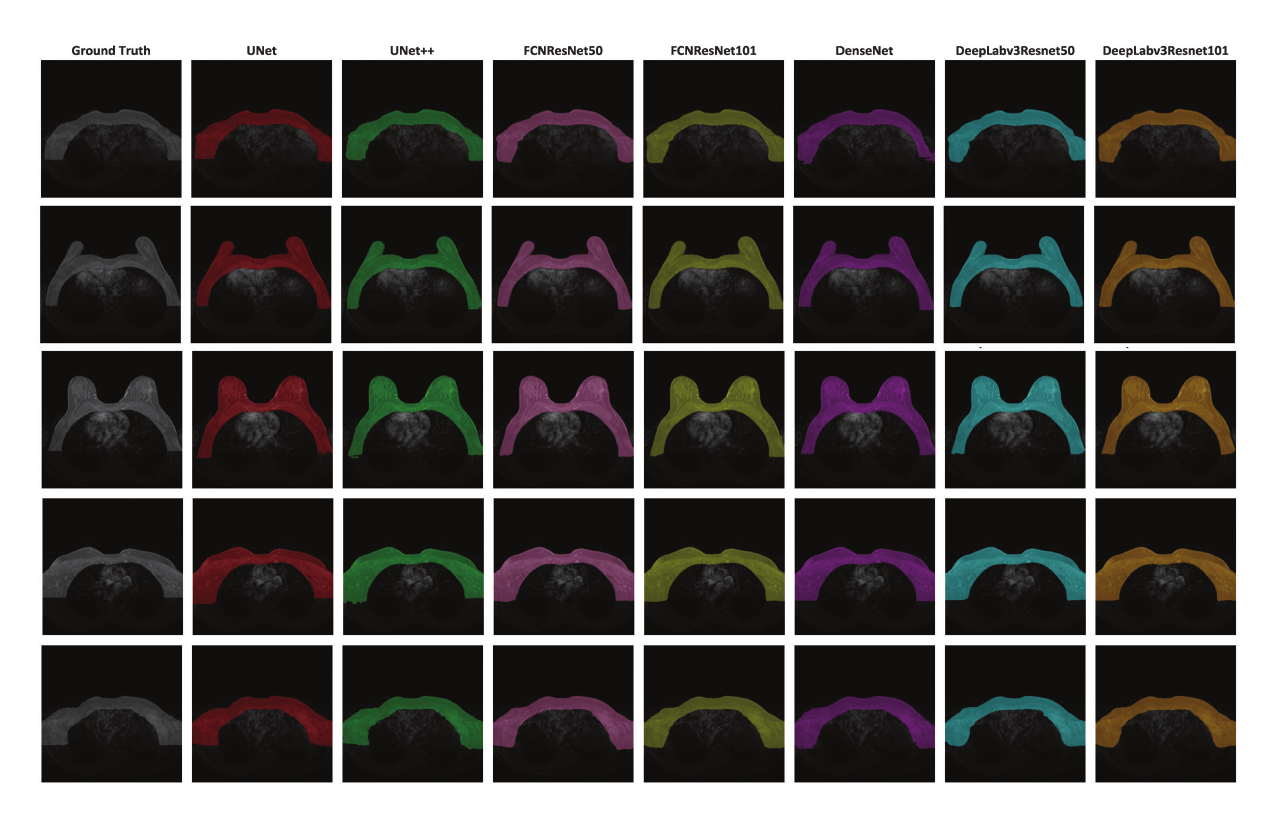}
\caption{\centering Model segmentation results across selected slices (from top to down row: first, 30\textsuperscript{th}, middle, 120\textsuperscript{th} and last slices)}\label{fig2}
\end{figure}

Figure \ref{fig2} illustrates the segmentation results of different models across selected slices, specifically the first, 30th, middle, 120th, and last slices. These slices were chosen to represent the progression from the initial to the final slices of the volume, allowing for a comprehensive comparison of each model's ability to maintain accuracy throughout the entire dataset.

\subsection{Internal breast segmentation and boundary detection}\label{sec3}
\begin{figure}[h!]
\centering
\includegraphics[width=0.8\textwidth]{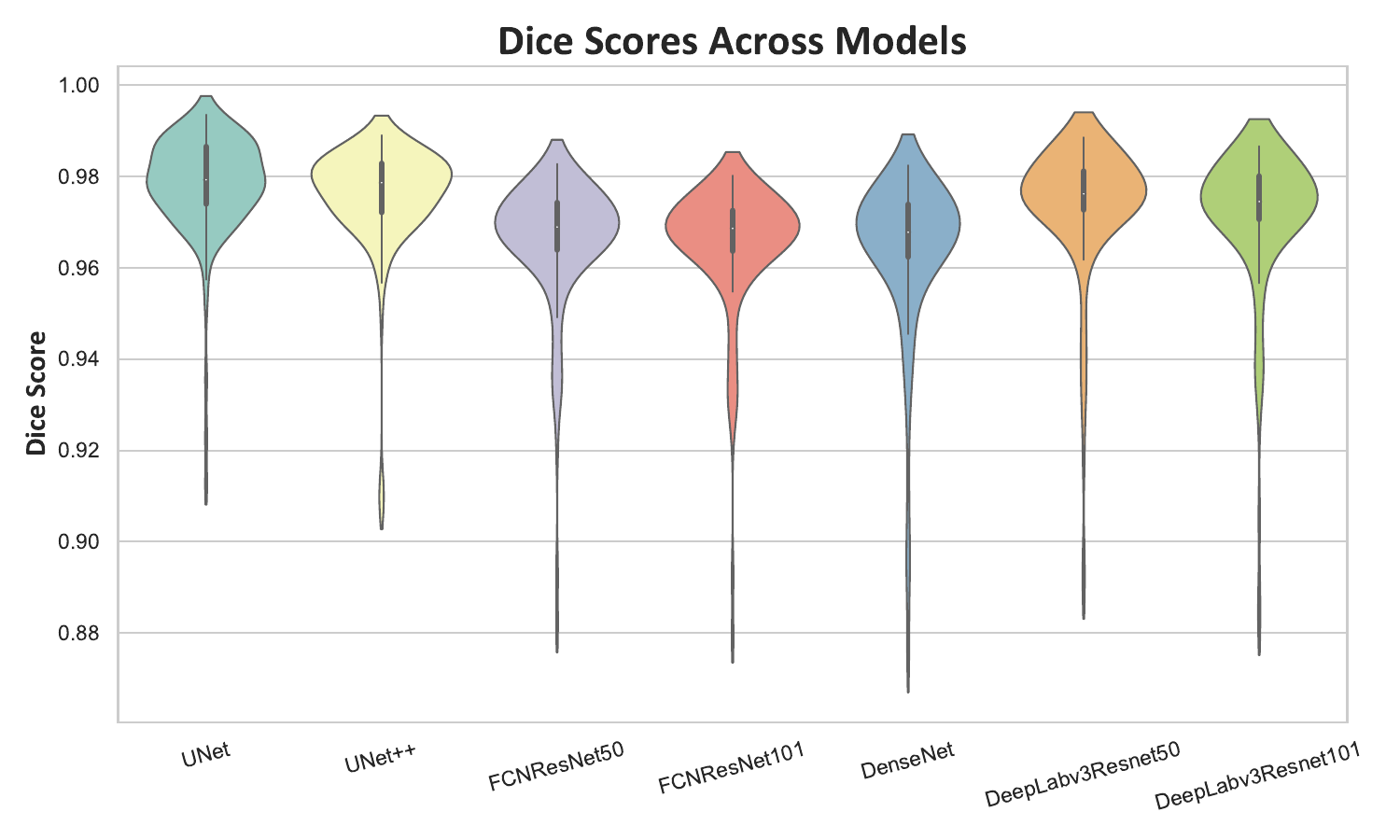}
\caption{\centering Dice score for different DL models on test subset}\label{fig3}
\end{figure}

Figure \ref{fig3} shows the distribution of Dice scores for each model on the test dataset, focusing on the median and range of performance. Notably, the UNet model has a median Dice score of 0.98, with a range from 0.91 to 0.995, highlighting its strong performance. Similarly, UNet++ achieves a median score of 0.98, with a range from 0.90 to 0.99. Close behind, DeepLabv3 with ResNet101 records a median of 0.975 with a slightly wider range from 0.88 to 0.99.
On the other hand, the FCN models with ResNet50 and ResNet101 backbones show median Dice scores of 0.970 and 0.972, respectively, with ranges approximately from 0.88 to 0.985 for both. In contrast, DenseNet, though having a similar median Dice score of 0.970, shows the widest range of 0.87 to 0.99, indicating more variability and less consistent segmentation accuracy.

\begin{figure}[h!]
\centering
\includegraphics[width=0.8\textwidth]{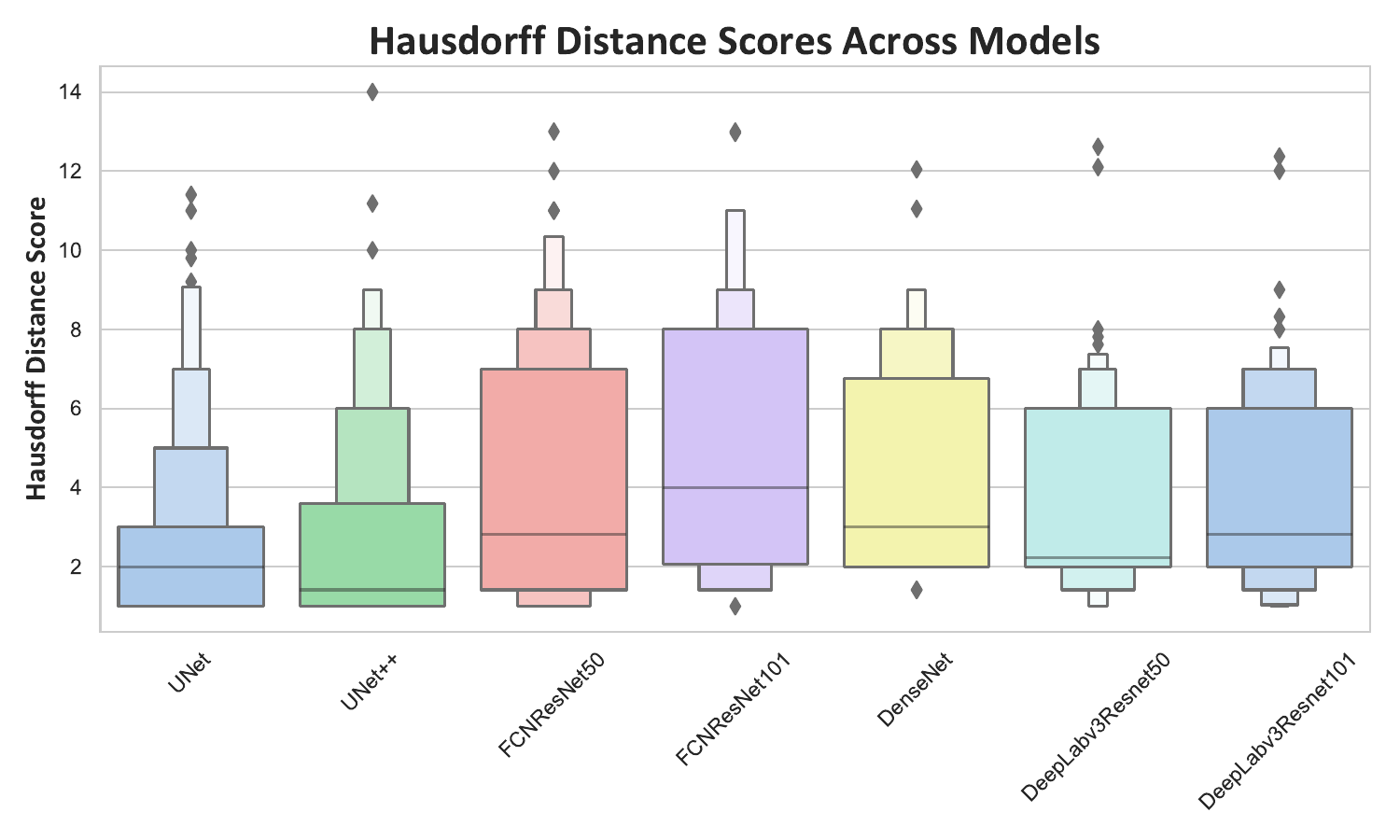}
\caption{\centering Hausdorff distance comparison across different models}\label{fig4}
\end{figure}

As was evident in Figure \ref{fig2}, the segmentation of the breast region was generally excellent across all models, but differences in boundary detection were observed. To further evaluate and compare the boundary detection capabilities of each model, Figure \ref{fig4} presents the Hausdorff distance on the test subset. As shown in the figure, UNet and UNet++ exhibit the lowest median Hausdorff distances, indicating their superior ability to accurately capture boundary details with minimal deviation. In contrast, the FCN models, particularly with ResNet50 and ResNet101 backbones, display higher median Hausdorff distances and a broader spread, highlighting greater variability in boundary detection and less precise segmentation at the edges.
Similarly, DenseNet shows a wide range of Hausdorff distances, with a higher number of outliers, indicating that while it may perform adequately in some cases, it struggles with boundary accuracy in others. On the other hand, DeepLabv3 with ResNet50 and ResNet101 demonstrate relatively lower median Hausdorff distances compared to FCN models, but with a few outliers, suggesting these models are generally reliable but may still occasionally falter in capturing fine boundary details.

\subsection{Training and inference time}\label{sec3}

Training and inference time are critical considerations in efficiency and cost of the modelling. Table \ref{tabel4} Demonstrates training time results of 10-fold, with mean and standard deviation, alongside average inference time in the test set for diverse architectures. As presented in the table, FCN with ResNet50 shows the shortest training time at 87 ± 18 minutes, while DenseNet exhibits the longest at 185 ± 56 minutes. In terms of inference time per slice, UNet performs the best with 126 milliseconds, whereas DenseNet requires significantly more time at 696 milliseconds, highlighting varying computational efficiencies across these models. Despite possessing fewer trained parameters, UNet++ required more time, 199 ± 21, to train each fold. On the other hand, DeepLabv3ResNet50 demonstrated superior performance, followed by FCNResNet50, with training times of 104 ± 22 minutes per fold.

\begin{table}[ht]
\centering
\caption{\centering Training and inference times across different models.}\label{tabel4}
\begin{tabular}{lcc}
\toprule
\textbf{Models} & \textbf{Training time per fold (min)} & \textbf{Inference time per slice (msec)} \\
\midrule
UNet & 136 $\pm$ 22 & 126 \\
UNet++ & 199 $\pm$ 21 & 152 \\
DenseNet & 185 $\pm$ 56 & 696 \\
FCNResNet50 & 87 $\pm$ 18 & 140 \\
FCNResNet101 & 149 $\pm$ 35 & 266 \\
DeepLabv3ResNet50 & 104 $\pm$ 22 & 161 \\
DeepLabv3ResNet101 & 178 $\pm$ 37 & 294 \\

\bottomrule
\end{tabular}
\label{tab:training_inference_times}
\end{table}

\begin{figure}[h!]
\centering
\includegraphics[width=1.1\textwidth]{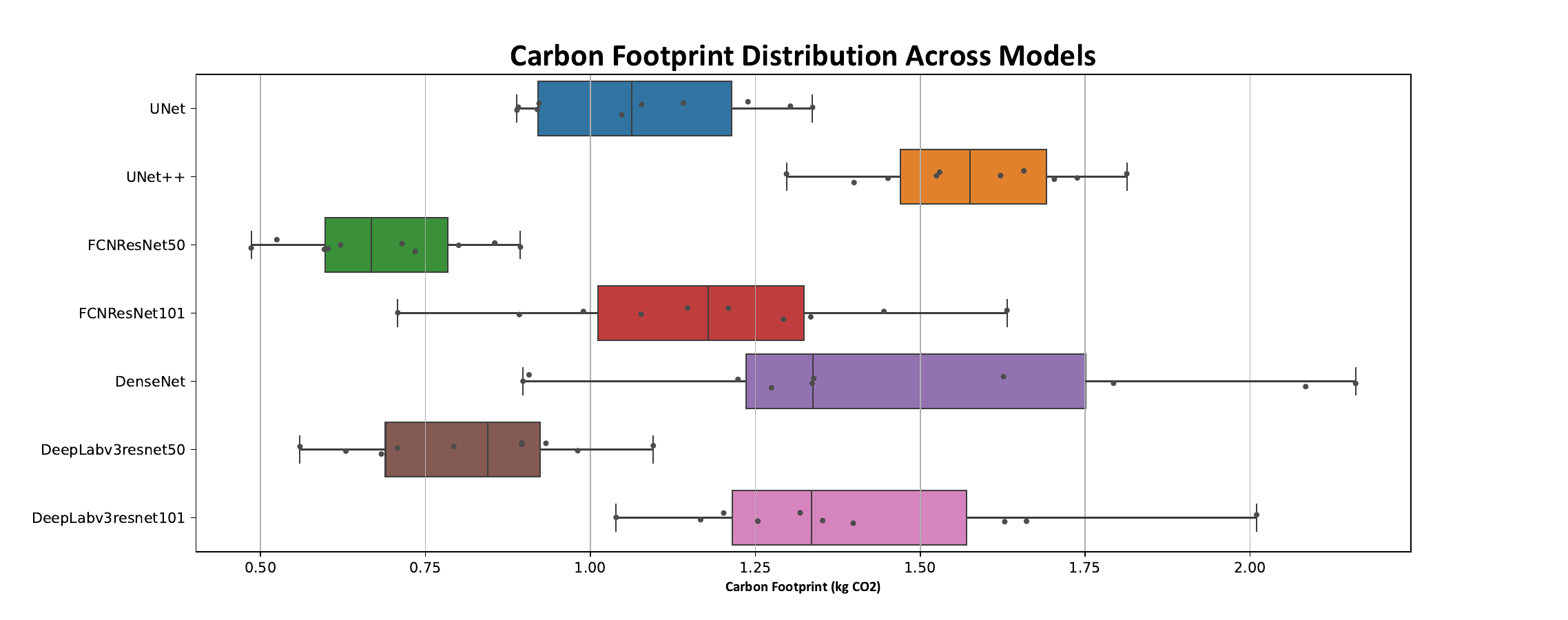}
\caption{\centering Carbon footprint across folds for various architectures}\label{fig5}
\end{figure}

Recent research underscores the growing importance of assessing the carbon footprint as a key factor in evaluating environmental sustainability across various models \cite{strubell2020energy,zhong2024impact,tamburrini2022ai}. Consequently, it is essential to investigate the carbon footprint associated with different network architectures during 10-fold cross-validation training. As illustrated in Figure \ref{fig5}, FCNResNet50 emerges as the most favorable and sustainable model, exhibiting a carbon footprint range of 0.45 to 0.85 kg CO2. This indicates that FCNResNet50 demonstrates the lowest environmental impact among the models analyzed. In contrast, DenseNet training is associated with higher energy consumption and, consequently, a larger carbon footprint. Following FCNResNet50, Deeplabv3ResNet50 displays the second-lowest carbon footprint, ranging from 0.55 to 1.15 kg CO2. Other models exhibit carbon footprints situated between these two extremes, reflecting a spectrum of environmental impacts.

\section{Discussion}\label{sec4}
This study offers a comprehensive evaluation of various state-of-the-art segmentation architectures applied to whole breast segmentation. The models examined include UNet, UNet++, DenseNet, FCNResNet50, FCNResNet101, DeepLabv3ResNet50, and DeepLabv3ResNet101. The results indicate significant differences in performance and training efficiency across these architectures, providing valuable insights into MR breast image analysis.

As outlined in Table \ref{tabel3}, UNet exhibited a higher Dice loss function during training compared to UNet++. This may be attributed to UNet’s simpler architecture, which could enable more effective generalization during validation. Conversely, UNet++ introduces added complexity through its nested and dense skip connections, which may contribute to slower convergence, as shown in Table \ref{tabel4}, or difficulties in optimizing all parameters effectively. This increased complexity may also render the model more susceptible to overfitting, as indicated by the higher variance in validation loss. The comparison between FCN with ResNet50 and ResNet101 underscores the impact of deeper networks. ResNet101, being a deeper model than ResNet50, generally allows for the learning of more complex features. However, the slightly higher training Dice loss in ResNet101 suggests that, while it has the potential to learn more detailed representations, the benefits may diminish, particularly if the dataset is insufficiently large to fully exploit the deeper network's capacity. Similarly, DeepLabv3 with ResNet50 and ResNet101 shows relatively close performance. The architecture of DeepLabv3, which incorporates atrous convolutions and multi-scale context aggregation, is intended to enhance feature extraction for semantic segmentation. The minor variation in Dice loss suggests that, although ResNet101 offers more layers and potentially improved feature extraction, the advantages are not significantly superior to those of ResNet50. This may indicate that the additional layers in ResNet101 are not fully utilized, or that the model's complexity poses challenges in training without overfitting. DenseNet demonstrates the highest Dice loss among the models. Although DenseNet’s architecture employs a dense connectivity pattern that encourages feature reuse and facilitates gradient flow during backpropagation, it lacks skip connections between the feature extractor and the expansion part of the network. Consequently, the poorest results can be attributed to this absence of skip connections.

Segmentation performance is illustrated in Figure \ref{fig2} for individual slices. In the first and last rows, corresponding to the first and last slices, significant differences are observed around the tails and the lower boundary, just above the heart and lungs. In contrast, the second and fourth rows, representing intermediate slices, exhibit improved boundary delineation across all models. The middle slice, depicted in the third row, demonstrates the best segmentation performance among all slices.

Segmentation efficiency can be evaluated based on two primary aspects: internal segmentation quality, measured by the Dice score, and accuracy in boundary delineation. For the assessment of boundary accuracy, the Hausdorff distance metric was employed to compare different models. Figure \ref{fig4} illustrates that UNet, UNet++, and DeepLabv3 with ResNets demonstrate superior performance in boundary detection. UNet and UNet++ excel in boundary detection due to their architectures, which are optimized for fine-grained segmentation tasks. DeepLabv3 with ResNets achieves precise boundary detection capabilities by leveraging the ASPP block integrated within its architecture.
FCNResNet50 emerges as the model with the smallest carbon footprint, indicating its relative efficiency in terms of energy consumption during training. This efficiency may be attributed to the balance between model complexity and the depth of the ResNet50 backbone, which is adequate to perform well without excessive computational demand. Similarly, DeepLabv3 with ResNet50 also exhibits a relatively low carbon footprint, owing to its effective use of ASPP and multi-scale context aggregation. In contrast, UNet++ demonstrates an unexpectedly high carbon footprint despite having fewer parameters than other models like DenseNet. This can be largely attributed to its complex architecture, which includes nested and dense skip connections.
In comparison to the study conducted in \cite{huo2021segmentation}, all architectures used in this study demonstrate superior results. Although the datasets differ, the influence of hyperparameters and preprocessing steps cannot be overlooked. Additionally , the MRI dataset used in this study, collected in 2008, remains relevant for breast region segmentation due to its imaging quality and detailed annotations, which still provide critical insights for developing and assessing segmentation models. This study offers a thorough examination of deep learning architectures specifically for breast region segmentation in DCE-MRI images. We explore the latest advancements in deep learning techniques and assess their applicability and performance in this crucial task. Future work could further enhance preprocessing methods and investigate various loss functions to evaluate additional effective factors influencing breast region segmentation.

\section{Conclusion}\label{sec5}
In this study, a deep learning (DL) pipeline was developed to evaluate performance of seven DL architectures for the segmentation of breast regions in DCE-MRI. A boundary was proposed for delineating breast borders, and manual annotation was carried out to provide accurate ground truth data. The efficiency of each architecture was assessed in terms of model training loss, training time, inference time, Dice score, boundary detection accuracy, and carbon footprint. The results indicated that UNet++ exhibited the best overall model performance, demonstrating superior accuracy in terms of Dice score. However, UNet showed better prediction accuracy on the validation set. On the other hand, FCNResNet50 emerged as the most efficient model concerning training time and carbon footprint, while UNet achieved the best inference time, making it suitable for real-time applications. These findings underscore the trade-offs between different models, highlighting that each architecture has its strengths and weaknesses depending on the evaluation criteria. For instance, UNet++ provides high segmentation accuracy but requires longer training times, whereas FCNResNet50 is more environmentally sustainable with a lower carbon footprint and quicker training times. This study emphasizes that the choice of model should be based on specific requirements and constraints of the application at hand. Given the increasing importance of sustainability, carbon footprint has become a crucial factor in model selection. Consequently, FCNResNet50 is identified as the most competitive model, balancing excellent performance and efficiency with a moderate inference time.

\section*{Declarations}
\subsection*{Ethics compliance}
All patients enrolled in the cohorts of the IMAGINE project received approval from the regional ethics committee.

\subsection*{Data availability}
The Stavanger dataset analyzed in this study contains sensitive patient information and therefore not publicly available. This ensures compliance with patient confidentiality and privacy regulations.

\subsection*{Code availability}
The code for data processing, analyzing and modelling is available on GitHub. To access the code repository, please follow the link on \href{https://github.com/SamNarimani-lab/Breast.git}{GitHub.}

\subsection*{Acknowledgment}
We would like to extend our sincere thanks to More and Romsdal Hospital Trust for their invaluable support and Stavanger University Hospital for providing resources and data. Special thanks go to Kathrine Røe Redalen, Iman Esmaili, Taraneh Ghasemi and Hadi Akbari Zadeh for their assistance throughout the research process. This work was funded by the Central Norway Regional Health Authority under the name of the IMAGINE project.

\subsection*{Author contributions}
\textbf{Sam Narimani}: Drafted the introduction, methods and materials, proposed a new boundary for breast region segmentation, and authored the results, discussion, and conclusion sections. Additionally, he contributed programming code. \\
\textbf{Solveig Roth Hoff}: Contributed to the writing of the introduction, the breast boundary definition, Discussion section, providing major insights.\\ 
\textbf{Kathinka Dæhli Kurz}: Responsible for data acquisition and preparation, and contributed to the discussion section. \\
\textbf{Kjell-Inge Gjesdal}: Set up MRI protocols. \\
\textbf{Jurgen Geisler}: Provided revisions and improvements to the introduction.\\
\textbf{Endre Grovik}: Supervised the project, acquired funding, and contributed to the revision of the introduction, methods and materials, results and discussion sections.



\begin{thebibliography}{32}
\ifx \bisbn   \undefined \def \bisbn  #1{ISBN #1}\fi
\ifx \binits  \undefined \def \binits#1{#1}\fi
\ifx \bauthor  \undefined \def \bauthor#1{#1}\fi
\ifx \batitle  \undefined \def \batitle#1{#1}\fi
\ifx \bjtitle  \undefined \def \bjtitle#1{#1}\fi
\ifx \bvolume  \undefined \def \bvolume#1{\textbf{#1}}\fi
\ifx \byear  \undefined \def \byear#1{#1}\fi
\ifx \bissue  \undefined \def \bissue#1{#1}\fi
\ifx \bfpage  \undefined \def \bfpage#1{#1}\fi
\ifx \blpage  \undefined \def \blpage #1{#1}\fi
\ifx \burl  \undefined \def \burl#1{\textsf{#1}}\fi
\ifx \doiurl  \undefined \def \doiurl#1{\url{https://doi.org/#1}}\fi
\ifx \betal  \undefined \def \betal{\textit{et al.}}\fi
\ifx \binstitute  \undefined \def \binstitute#1{#1}\fi
\ifx \binstitutionaled  \undefined \def \binstitutionaled#1{#1}\fi
\ifx \bctitle  \undefined \def \bctitle#1{#1}\fi
\ifx \beditor  \undefined \def \beditor#1{#1}\fi
\ifx \bpublisher  \undefined \def \bpublisher#1{#1}\fi
\ifx \bbtitle  \undefined \def \bbtitle#1{#1}\fi
\ifx \bedition  \undefined \def \bedition#1{#1}\fi
\ifx \bseriesno  \undefined \def \bseriesno#1{#1}\fi
\ifx \blocation  \undefined \def \blocation#1{#1}\fi
\ifx \bsertitle  \undefined \def \bsertitle#1{#1}\fi
\ifx \bsnm \undefined \def \bsnm#1{#1}\fi
\ifx \bsuffix \undefined \def \bsuffix#1{#1}\fi
\ifx \bparticle \undefined \def \bparticle#1{#1}\fi
\ifx \barticle \undefined \def \barticle#1{#1}\fi
\bibcommenthead
\ifx \bconfdate \undefined \def \bconfdate #1{#1}\fi
\ifx \botherref \undefined \def \botherref #1{#1}\fi
\ifx \url \undefined \def \url#1{\textsf{#1}}\fi
\ifx \bchapter \undefined \def \bchapter#1{#1}\fi
\ifx \bbook \undefined \def \bbook#1{#1}\fi
\ifx \bcomment \undefined \def \bcomment#1{#1}\fi
\ifx \oauthor \undefined \def \oauthor#1{#1}\fi
\ifx \citeauthoryear \undefined \def \citeauthoryear#1{#1}\fi
\ifx \endbibitem  \undefined \def \endbibitem {}\fi
\ifx \bconflocation  \undefined \def \bconflocation#1{#1}\fi
\ifx \arxivurl  \undefined \def \arxivurl#1{\textsf{#1}}\fi
\csname PreBibitemsHook\endcsname

\bibitem[\protect\citeauthoryear{Acciavatti et~al.}{2023}]{acciavatti2023beyond}
\begin{barticle}
\bauthor{\bsnm{Acciavatti}, \binits{R.J.}},
\bauthor{\bsnm{Lee}, \binits{S.H.}},
\bauthor{\bsnm{Reig}, \binits{B.}},
\bauthor{\bsnm{Moy}, \binits{L.}},
\bauthor{\bsnm{Conant}, \binits{E.F.}},
\bauthor{\bsnm{Kontos}, \binits{D.}},
\bauthor{\bsnm{Moon}, \binits{W.K.}}:
\batitle{Beyond breast density: risk measures for breast cancer in multiple imaging modalities}.
\bjtitle{Radiology}
\bvolume{306}(\bissue{3}),
\bfpage{222575}
(\byear{2023})
\end{barticle}
\endbibitem

\bibitem[\protect\citeauthoryear{Lew et~al.}{2024}]{lew2024publicly}
\begin{barticle}
\bauthor{\bsnm{Lew}, \binits{C.O.}},
\bauthor{\bsnm{Harouni}, \binits{M.}},
\bauthor{\bsnm{Kirksey}, \binits{E.R.}},
\bauthor{\bsnm{Kang}, \binits{E.J.}},
\bauthor{\bsnm{Dong}, \binits{H.}},
\bauthor{\bsnm{Gu}, \binits{H.}},
\bauthor{\bsnm{Grimm}, \binits{L.J.}},
\bauthor{\bsnm{Walsh}, \binits{R.}},
\bauthor{\bsnm{Lowell}, \binits{D.A.}},
\bauthor{\bsnm{Mazurowski}, \binits{M.A.}}:
\batitle{A publicly available deep learning model and dataset for segmentation of breast, fibroglandular tissue, and vessels in breast mri}.
\bjtitle{Scientific reports}
\bvolume{14}(\bissue{1}),
\bfpage{5383}
(\byear{2024})
\end{barticle}
\endbibitem

\bibitem[\protect\citeauthoryear{Ojala et~al.}{2001}]{ojala2001accurate}
\begin{barticle}
\bauthor{\bsnm{Ojala}, \binits{T.}},
\bauthor{\bsnm{N{\"a}ppi}, \binits{J.}},
\bauthor{\bsnm{Nevalainen}, \binits{O.}}:
\batitle{Accurate segmentation of the breast region from digitized mammograms}.
\bjtitle{Computerized medical imaging and graphics}
\bvolume{25}(\bissue{1}),
\bfpage{47}--\blpage{59}
(\byear{2001})
\end{barticle}
\endbibitem

\bibitem[\protect\citeauthoryear{Jiao et~al.}{2020}]{jiao2020deep}
\begin{barticle}
\bauthor{\bsnm{Jiao}, \binits{H.}},
\bauthor{\bsnm{Jiang}, \binits{X.}},
\bauthor{\bsnm{Pang}, \binits{Z.}},
\bauthor{\bsnm{Lin}, \binits{X.}},
\bauthor{\bsnm{Huang}, \binits{Y.}},
\bauthor{\bsnm{Li}, \binits{L.}}:
\batitle{Deep convolutional neural networks-based automatic breast segmentation and mass detection in dce-mri}.
\bjtitle{Computational and Mathematical Methods in Medicine}
\bvolume{2020}(\bissue{1}),
\bfpage{2413706}
(\byear{2020})
\end{barticle}
\endbibitem

\bibitem[\protect\citeauthoryear{Ivanovska et~al.}{2019}]{ivanovska2019deep}
\begin{barticle}
\bauthor{\bsnm{Ivanovska}, \binits{T.}},
\bauthor{\bsnm{Jentschke}, \binits{T.G.}},
\bauthor{\bsnm{Daboul}, \binits{A.}},
\bauthor{\bsnm{Hegenscheid}, \binits{K.}},
\bauthor{\bsnm{V{\"o}lzke}, \binits{H.}},
\bauthor{\bsnm{W{\"o}rg{\"o}tter}, \binits{F.}}:
\batitle{A deep learning framework for efficient analysis of breast volume and fibroglandular tissue using mr data with strong artifacts}.
\bjtitle{International journal of computer assisted radiology and surgery}
\bvolume{14},
\bfpage{1627}--\blpage{1633}
(\byear{2019})
\end{barticle}
\endbibitem

\bibitem[\protect\citeauthoryear{Huo et~al.}{2021}]{huo2021segmentation}
\begin{barticle}
\bauthor{\bsnm{Huo}, \binits{L.}},
\bauthor{\bsnm{Hu}, \binits{X.}},
\bauthor{\bsnm{Xiao}, \binits{Q.}},
\bauthor{\bsnm{Gu}, \binits{Y.}},
\bauthor{\bsnm{Chu}, \binits{X.}},
\bauthor{\bsnm{Jiang}, \binits{L.}}:
\batitle{Segmentation of whole breast and fibroglandular tissue using nnu-net in dynamic contrast enhanced mr images}.
\bjtitle{Magnetic Resonance Imaging}
\bvolume{82},
\bfpage{31}--\blpage{41}
(\byear{2021})
\end{barticle}
\endbibitem

\bibitem[\protect\citeauthoryear{van~der Velden et~al.}{2020}]{van2020volumetric}
\begin{barticle}
\bauthor{\bsnm{Velden}, \binits{B.H.}},
\bauthor{\bsnm{Janse}, \binits{M.H.}},
\bauthor{\bsnm{Ragusi}, \binits{M.A.}},
\bauthor{\bsnm{Loo}, \binits{C.E.}},
\bauthor{\bsnm{Gilhuijs}, \binits{K.G.}}:
\batitle{Volumetric breast density estimation on mri using explainable deep learning regression}.
\bjtitle{Scientific reports}
\bvolume{10}(\bissue{1}),
\bfpage{18095}
(\byear{2020})
\end{barticle}
\endbibitem

\bibitem[\protect\citeauthoryear{Yue et~al.}{2022}]{yue2022deep}
\begin{barticle}
\bauthor{\bsnm{Yue}, \binits{W.}},
\bauthor{\bsnm{Zhang}, \binits{H.}},
\bauthor{\bsnm{Zhou}, \binits{J.}},
\bauthor{\bsnm{Li}, \binits{G.}},
\bauthor{\bsnm{Tang}, \binits{Z.}},
\bauthor{\bsnm{Sun}, \binits{Z.}},
\bauthor{\bsnm{Cai}, \binits{J.}},
\bauthor{\bsnm{Tian}, \binits{N.}},
\bauthor{\bsnm{Gao}, \binits{S.}},
\bauthor{\bsnm{Dong}, \binits{J.}}, \betal:
\batitle{Deep learning-based automatic segmentation for size and volumetric measurement of breast cancer on magnetic resonance imaging}.
\bjtitle{Frontiers in Oncology}
\bvolume{12},
\bfpage{984626}
(\byear{2022})
\end{barticle}
\endbibitem

\bibitem[\protect\citeauthoryear{Giannini et~al.}{2010}]{giannini2010fully}
\begin{bchapter}
\bauthor{\bsnm{Giannini}, \binits{V.}},
\bauthor{\bsnm{Vignati}, \binits{A.}},
\bauthor{\bsnm{Morra}, \binits{L.}},
\bauthor{\bsnm{Persano}, \binits{D.}},
\bauthor{\bsnm{Brizzi}, \binits{D.}},
\bauthor{\bsnm{Carbonaro}, \binits{L.}},
\bauthor{\bsnm{Bert}, \binits{A.}},
\bauthor{\bsnm{Sardanelli}, \binits{F.}},
\bauthor{\bsnm{Regge}, \binits{D.}}:
\bctitle{A fully automatic algorithm for segmentation of the breasts in dce-mr images}.
In: \bbtitle{2010 Annual International Conference of the IEEE Engineering in Medicine and Biology},
pp. \bfpage{3146}--\blpage{3149}
(\byear{2010}).
\bcomment{IEEE}
\end{bchapter}
\endbibitem

\bibitem[\protect\citeauthoryear{Saffari et~al.}{2020}]{saffari2020fully}
\begin{barticle}
\bauthor{\bsnm{Saffari}, \binits{N.}},
\bauthor{\bsnm{Rashwan}, \binits{H.A.}},
\bauthor{\bsnm{Abdel-Nasser}, \binits{M.}},
\bauthor{\bsnm{Kumar~Singh}, \binits{V.}},
\bauthor{\bsnm{Arenas}, \binits{M.}},
\bauthor{\bsnm{Mangina}, \binits{E.}},
\bauthor{\bsnm{Herrera}, \binits{B.}},
\bauthor{\bsnm{Puig}, \binits{D.}}:
\batitle{Fully automated breast density segmentation and classification using deep learning}.
\bjtitle{Diagnostics}
\bvolume{10}(\bissue{11}),
\bfpage{988}
(\byear{2020})
\end{barticle}
\endbibitem

\bibitem[\protect\citeauthoryear{Erta{\c{s}} et~al.}{2008}]{ertacs2008breast}
\begin{barticle}
\bauthor{\bsnm{Erta{\c{s}}}, \binits{G.}},
\bauthor{\bsnm{G{\"u}l{\c{c}}{\"u}r}, \binits{H.{\"O}.}},
\bauthor{\bsnm{Osman}, \binits{O.}},
\bauthor{\bsnm{U{\c{c}}an}, \binits{O.N.}},
\bauthor{\bsnm{Tunac{\i}}, \binits{M.}},
\bauthor{\bsnm{Dursun}, \binits{M.}}:
\batitle{Breast mr segmentation and lesion detection with cellular neural networks and 3d template matching}.
\bjtitle{Computers in biology and medicine}
\bvolume{38}(\bissue{1}),
\bfpage{116}--\blpage{126}
(\byear{2008})
\end{barticle}
\endbibitem

\bibitem[\protect\citeauthoryear{Zhang et~al.}{2019}]{zhang2019automatic}
\begin{barticle}
\bauthor{\bsnm{Zhang}, \binits{Y.}},
\bauthor{\bsnm{Chen}, \binits{J.-H.}},
\bauthor{\bsnm{Chang}, \binits{K.-T.}},
\bauthor{\bsnm{Park}, \binits{V.Y.}},
\bauthor{\bsnm{Kim}, \binits{M.J.}},
\bauthor{\bsnm{Chan}, \binits{S.}},
\bauthor{\bsnm{Chang}, \binits{P.}},
\bauthor{\bsnm{Chow}, \binits{D.}},
\bauthor{\bsnm{Luk}, \binits{A.}},
\bauthor{\bsnm{Kwong}, \binits{T.}}, \betal:
\batitle{Automatic breast and fibroglandular tissue segmentation in breast mri using deep learning by a fully-convolutional residual neural network u-net}.
\bjtitle{Academic radiology}
\bvolume{26}(\bissue{11}),
\bfpage{1526}--\blpage{1535}
(\byear{2019})
\end{barticle}
\endbibitem

\bibitem[\protect\citeauthoryear{Piantadosi et~al.}{2018}]{piantadosi2018breast}
\begin{bchapter}
\bauthor{\bsnm{Piantadosi}, \binits{G.}},
\bauthor{\bsnm{Sansone}, \binits{M.}},
\bauthor{\bsnm{Sansone}, \binits{C.}}:
\bctitle{Breast segmentation in mri via u-net deep convolutional neural networks}.
In: \bbtitle{2018 24th International Conference on Pattern Recognition (ICPR)},
pp. \bfpage{3917}--\blpage{3922}
(\byear{2018}).
\bcomment{IEEE}
\end{bchapter}
\endbibitem

\bibitem[\protect\citeauthoryear{Xu et~al.}{2018}]{xu2018breast}
\begin{bchapter}
\bauthor{\bsnm{Xu}, \binits{X.}},
\bauthor{\bsnm{Fu}, \binits{L.}},
\bauthor{\bsnm{Chen}, \binits{Y.}},
\bauthor{\bsnm{Larsson}, \binits{R.}},
\bauthor{\bsnm{Zhang}, \binits{D.}},
\bauthor{\bsnm{Suo}, \binits{S.}},
\bauthor{\bsnm{Hua}, \binits{J.}},
\bauthor{\bsnm{Zhao}, \binits{J.}}:
\bctitle{Breast region segmentation being convolutional neural network in dynamic contrast enhanced mri}.
In: \bbtitle{2018 40th Annual International Conference of the IEEE Engineering in Medicine and Biology Society (EMBC)},
pp. \bfpage{750}--\blpage{753}
(\byear{2018}).
\bcomment{IEEE}
\end{bchapter}
\endbibitem

\bibitem[\protect\citeauthoryear{Yao et~al.}{2009}]{yao2009breast}
\begin{barticle}
\bauthor{\bsnm{Yao}, \binits{J.}},
\bauthor{\bsnm{Chen}, \binits{J.}},
\bauthor{\bsnm{Chow}, \binits{C.}}:
\batitle{Breast tumor analysis in dynamic contrast enhanced mri using texture features and wavelet transform}.
\bjtitle{IEEE Journal of selected topics in signal processing}
\bvolume{3}(\bissue{1}),
\bfpage{94}--\blpage{100}
(\byear{2009})
\end{barticle}
\endbibitem

\bibitem[\protect\citeauthoryear{Nie et~al.}{2008}]{nie2008development}
\begin{barticle}
\bauthor{\bsnm{Nie}, \binits{K.}},
\bauthor{\bsnm{Chen}, \binits{J.-H.}},
\bauthor{\bsnm{Chan}, \binits{S.}},
\bauthor{\bsnm{Chau}, \binits{M.-K.I.}},
\bauthor{\bsnm{Yu}, \binits{H.J.}},
\bauthor{\bsnm{Bahri}, \binits{S.}},
\bauthor{\bsnm{Tseng}, \binits{T.}},
\bauthor{\bsnm{Nalcioglu}, \binits{O.}},
\bauthor{\bsnm{Su}, \binits{M.-Y.}}:
\batitle{Development of a quantitative method for analysis of breast density based on three-dimensional breast mri}.
\bjtitle{Medical physics}
\bvolume{35}(\bissue{12}),
\bfpage{5253}--\blpage{5262}
(\byear{2008})
\end{barticle}
\endbibitem

\bibitem[\protect\citeauthoryear{Sui et~al.}{2021}]{sui2021breast}
\begin{bchapter}
\bauthor{\bsnm{Sui}, \binits{D.}},
\bauthor{\bsnm{Huang}, \binits{Z.}},
\bauthor{\bsnm{Song}, \binits{X.}},
\bauthor{\bsnm{Zhang}, \binits{Y.}},
\bauthor{\bsnm{Wang}, \binits{Y.}},
\bauthor{\bsnm{Zhang}, \binits{L.}}:
\bctitle{Breast regions segmentation based on u-net++ from dce-mri image sequences}.
In: \bbtitle{Journal of Physics: Conference Series},
vol. \bseriesno{1748},
p. \bfpage{042058}
(\byear{2021}).
\bcomment{IOP Publishing}
\end{bchapter}
\endbibitem

\bibitem[\protect\citeauthoryear{Radak et~al.}{2023}]{radak2023machine}
\begin{barticle}
\bauthor{\bsnm{Radak}, \binits{M.}},
\bauthor{\bsnm{Lafta}, \binits{H.Y.}},
\bauthor{\bsnm{Fallahi}, \binits{H.}}:
\batitle{Machine learning and deep learning techniques for breast cancer diagnosis and classification: a comprehensive review of medical imaging studies}.
\bjtitle{Journal of Cancer Research and Clinical Oncology}
\bvolume{149}(\bissue{12}),
\bfpage{10473}--\blpage{10491}
(\byear{2023})
\end{barticle}
\endbibitem

\bibitem[\protect\citeauthoryear{Ronneberger et~al.}{2015}]{ronneberger2015u}
\begin{bchapter}
\bauthor{\bsnm{Ronneberger}, \binits{O.}},
\bauthor{\bsnm{Fischer}, \binits{P.}},
\bauthor{\bsnm{Brox}, \binits{T.}}:
\bctitle{U-net: Convolutional networks for biomedical image segmentation}.
In: \bbtitle{Medical Image Computing and Computer-assisted intervention--MICCAI 2015: 18th International Conference, Munich, Germany, October 5-9, 2015, Proceedings, Part III 18},
pp. \bfpage{234}--\blpage{241}
(\byear{2015}).
\bcomment{Springer}
\end{bchapter}
\endbibitem

\bibitem[\protect\citeauthoryear{Zhou et~al.}{2018}]{zhou2018unet++}
\begin{bchapter}
\bauthor{\bsnm{Zhou}, \binits{Z.}},
\bauthor{\bsnm{Rahman~Siddiquee}, \binits{M.M.}},
\bauthor{\bsnm{Tajbakhsh}, \binits{N.}},
\bauthor{\bsnm{Liang}, \binits{J.}}:
\bctitle{Unet++: A nested u-net architecture for medical image segmentation}.
In: \bbtitle{Deep Learning in Medical Image Analysis and Multimodal Learning for Clinical Decision Support: 4th International Workshop, DLMIA 2018, and 8th International Workshop, ML-CDS 2018, Held in Conjunction with MICCAI 2018, Granada, Spain, September 20, 2018, Proceedings 4},
pp. \bfpage{3}--\blpage{11}
(\byear{2018}).
\bcomment{Springer}
\end{bchapter}
\endbibitem

\bibitem[\protect\citeauthoryear{Huang et~al.}{2017}]{huang2017densely}
\begin{bchapter}
\bauthor{\bsnm{Huang}, \binits{G.}},
\bauthor{\bsnm{Liu}, \binits{Z.}},
\bauthor{\bsnm{Van Der~Maaten}, \binits{L.}},
\bauthor{\bsnm{Weinberger}, \binits{K.Q.}}:
\bctitle{Densely connected convolutional networks}.
In: \bbtitle{Proceedings of the IEEE Conference on Computer Vision and Pattern Recognition},
pp. \bfpage{4700}--\blpage{4708}
(\byear{2017})
\end{bchapter}
\endbibitem

\bibitem[\protect\citeauthoryear{Long et~al.}{2015}]{long2015fully}
\begin{bchapter}
\bauthor{\bsnm{Long}, \binits{J.}},
\bauthor{\bsnm{Shelhamer}, \binits{E.}},
\bauthor{\bsnm{Darrell}, \binits{T.}}:
\bctitle{Fully convolutional networks for semantic segmentation}.
In: \bbtitle{Proceedings of the IEEE Conference on Computer Vision and Pattern Recognition},
pp. \bfpage{3431}--\blpage{3440}
(\byear{2015})
\end{bchapter}
\endbibitem

\bibitem[\protect\citeauthoryear{He et~al.}{2016}]{he2016deep}
\begin{bchapter}
\bauthor{\bsnm{He}, \binits{K.}},
\bauthor{\bsnm{Zhang}, \binits{X.}},
\bauthor{\bsnm{Ren}, \binits{S.}},
\bauthor{\bsnm{Sun}, \binits{J.}}:
\bctitle{Deep residual learning for image recognition}.
In: \bbtitle{Proceedings of the IEEE Conference on Computer Vision and Pattern Recognition},
pp. \bfpage{770}--\blpage{778}
(\byear{2016})
\end{bchapter}
\endbibitem

\bibitem[\protect\citeauthoryear{Chen}{2017}]{chen2017rethinking}
\begin{botherref}
\oauthor{\bsnm{Chen}, \binits{L.-C.}}:
Rethinking atrous convolution for semantic image segmentation.
arXiv preprint arXiv:1706.05587
(2017)
\end{botherref}
\endbibitem

\bibitem[\protect\citeauthoryear{Dice}{1945}]{dice1945measures}
\begin{barticle}
\bauthor{\bsnm{Dice}, \binits{L.R.}}:
\batitle{Measures of the amount of ecologic association between species}.
\bjtitle{Ecology}
\bvolume{26}(\bissue{3}),
\bfpage{297}--\blpage{302}
(\byear{1945})
\end{barticle}
\endbibitem

\bibitem[\protect\citeauthoryear{Kohavi}{1995}]{kohavi1995study}
\begin{botherref}
\oauthor{\bsnm{Kohavi}, \binits{R.}}:
A study of cross-validation and bootstrap for accuracy estimation and model selection.
Morgan Kaufman Publishing
(1995)
\end{botherref}
\endbibitem

\bibitem[\protect\citeauthoryear{Rosado-Toro et~al.}{2015}]{rosado2015automated}
\begin{barticle}
\bauthor{\bsnm{Rosado-Toro}, \binits{J.A.}},
\bauthor{\bsnm{Barr}, \binits{T.}},
\bauthor{\bsnm{Galons}, \binits{J.-P.}},
\bauthor{\bsnm{Marron}, \binits{M.T.}},
\bauthor{\bsnm{Stopeck}, \binits{A.}},
\bauthor{\bsnm{Thomson}, \binits{C.}},
\bauthor{\bsnm{Thompson}, \binits{P.}},
\bauthor{\bsnm{Carroll}, \binits{D.}},
\bauthor{\bsnm{Wolf}, \binits{E.}},
\bauthor{\bsnm{Altbach}, \binits{M.I.}}, \betal:
\batitle{Automated breast segmentation of fat and water mr images using dynamic programming}.
\bjtitle{Academic radiology}
\bvolume{22}(\bissue{2}),
\bfpage{139}--\blpage{148}
(\byear{2015})
\end{barticle}
\endbibitem

\bibitem[\protect\citeauthoryear{Fooladivanda et~al.}{2017}]{fooladivanda2017breast}
\begin{barticle}
\bauthor{\bsnm{Fooladivanda}, \binits{A.}},
\bauthor{\bsnm{Shokouhi}, \binits{S.B.}},
\bauthor{\bsnm{Ahmadinejad}, \binits{N.}}:
\batitle{Breast-region segmentation in mri using chest region atlas and svm}.
\bjtitle{Turkish Journal of Electrical Engineering and Computer Sciences}
\bvolume{25}(\bissue{6}),
\bfpage{4575}--\blpage{4592}
(\byear{2017})
\end{barticle}
\endbibitem

\bibitem[\protect\citeauthoryear{{NVIDIA Corporation}}{2024}]{nvidia4090}
\begin{botherref}
\oauthor{\bsnm{{NVIDIA Corporation}}}:
{GeForce RTX 4090 Graphics Card}.
\url{https://www.nvidia.com/nb-no/geforce/graphics-cards/40-series/rtx-4090/}.
Accessed: 2024-09-02
(2024)
\end{botherref}
\endbibitem

\bibitem[\protect\citeauthoryear{Strubell et~al.}{2020}]{strubell2020energy}
\begin{bchapter}
\bauthor{\bsnm{Strubell}, \binits{E.}},
\bauthor{\bsnm{Ganesh}, \binits{A.}},
\bauthor{\bsnm{McCallum}, \binits{A.}}:
\bctitle{Energy and policy considerations for modern deep learning research}.
In: \bbtitle{Proceedings of the AAAI Conference on Artificial Intelligence},
vol. \bseriesno{34},
pp. \bfpage{13693}--\blpage{13696}
(\byear{2020})
\end{bchapter}
\endbibitem

\bibitem[\protect\citeauthoryear{Zhong et~al.}{2024}]{zhong2024impact}
\begin{barticle}
\bauthor{\bsnm{Zhong}, \binits{J.}},
\bauthor{\bsnm{Zhong}, \binits{Y.}},
\bauthor{\bsnm{Han}, \binits{M.}},
\bauthor{\bsnm{Yang}, \binits{T.}},
\bauthor{\bsnm{Zhang}, \binits{Q.}}:
\batitle{The impact of ai on carbon emissions: evidence from 66 countries}.
\bjtitle{Applied Economics}
\bvolume{56}(\bissue{25}),
\bfpage{2975}--\blpage{2989}
(\byear{2024})
\end{barticle}
\endbibitem

\bibitem[\protect\citeauthoryear{Tamburrini}{2022}]{tamburrini2022ai}
\begin{barticle}
\bauthor{\bsnm{Tamburrini}, \binits{G.}}:
\batitle{The ai carbon footprint and responsibilities of ai scientists}.
\bjtitle{Philosophies}
\bvolume{7}(\bissue{1}),
\bfpage{4}
(\byear{2022})
\end{barticle}
\endbibitem

\end{thebibliography}

\end{document}